\documentclass[
 amsmath,amssymb,
 aps,
 prx,
11pt, twocolumn]{revtex4-2}
\usepackage{setspace}
\usepackage{natbib}

\usepackage[pdftex]{graphicx}
\pdfoutput=1
\usepackage{dcolumn}
\usepackage{bm}

	\usepackage{color}
\usepackage[usenames,dvipsnames,svgnames,table]{xcolor}
\usepackage[colorlinks=true,
            linkcolor=blue,
            citecolor=blue]{hyperref}

\makeatletter \renewcommand\@make@capt@title[2]{%
\@ifx@empty\float@link{\@firstofone}{\expandafter\href\expandafter{\float@link}}%
\sffamily{\textbf{#1}}\@caption@fignum@sep#2 }

\begin{document}

\preprint{APS/123-QED}
\title{\color{black}\LARGE Photoinduced Band Renormalization Effects in ZrSiS Topological Nodal-line Semimetal  }

\author{Somnath Biswas$^{a}$}

 \author{Ioannis Petrides$^{b}$}
 \author{Robert J. Kirby$^{a}$}
 \author{Catrina Oberg$^{a}$}
 \author{Sebastian Klemenz$^{a}$}
 \author{Caroline Weinberg$^{a}$}
 \author{Austin Ferrenti$^{a}$}
 \author{Prineha Narang$^{b}$}
  \email{prineha@seas.harvard.edu}
 \author{Leslie Schoop$^{a}$}%
 \email{lschoop@princeton.edu}
\author{Gregory D. Scholes$^{a}$}%
 \email{gscholes@princeton.edu}
\affiliation{$^{a}$
 Department of Chemistry, Princeton University, Princeton, New Jersey 08544, USA
}%
 \affiliation{$^{b}$ John A. Paulson School of Engineering and Applied Sciences, Harvard University, Cambridge, Massachusetts 02138, USA}%

\begin{abstract}
\noindent Out-of-equilibrium effects provide an elegant pathway to probing and understanding the underlying physics of topological materials. 
Creating exotic states of matter using ultrafast optical pulses in particular has shown promise towards controlling electronic band structure properties.
Of recent interest is band renormalization in Dirac and Weyl semimetals as it leads to direct physical observables through the enhancement of the effective mass, or, in the shift of resonant energies.
Here we provide experimental and theoretical signatures of photo-induced renormalization of the electronic band structure in a topological nodal line semimetal ZrSiS. 
Specifically, we show how the change of the transient reflectivity spectra under femtosecond optical excitations is induced by out-of-equilibrium effects that renormalize the kinetic energy of electrons. 
We associate the observed spectral shift to an enhancement of the effective mass and to a red-shift of the resonant frequency as a function of pump field strength.
Finally, we show that the transient relaxation dynamics of the reflectivity is primarily an electronic effect with negligible phononic contribution.
Our study presents the modifications of electronic properties in ZrSiS using ultrashort pulses, and demonstrates the potential of this approach in creating photo-induced phases in topological quantum mater through an all-optical route.

\end{abstract}

\maketitle

\section{Introduction}
Dirac nodal-line semimetals (NLSMs) of the family ZrSiX, with X = S, Se, or Te, have been the subject of intense research due to the presence of topologically nontrivial states~ \cite{schoop2016dirac,neupane2016observation}, unusually large magnetoresistance~\cite{wang2016evidence,ali2016butterfly,singha2017large}, high carrier mobility~\cite{matusiak2017thermoelectric,sankar2017crystal}, strong Zeeman splitting~\cite{hu2017nearly}, and frequency independent optical conductivity~\cite{schilling2017flat}. 
In addition to these remarkable properties, the presence of extended linear band crossings along lines/loops in the Brillouin zone of NLSMs are anticipated to show enhanced correlation effects~\cite{liu2017correlation,armitage2018weyl,munoz2020many,shao2020electronic}. 
In fact, quantum oscillations studies of ZrSiS in high external magnetic fields showed an enhancement of the effective mass of the low-energy quasiparticles~\cite{pezzini2018unconventional}. 
The ability to manipulate the electronic properties in such material is therefore a primary driving force towards understanding and discovering new topological phases of matter.~\cite{narang2021topology,juraschek2021magnetic,disa2021engineering}


 \begin{figure*}[ht]
\includegraphics[width=6.3in]
{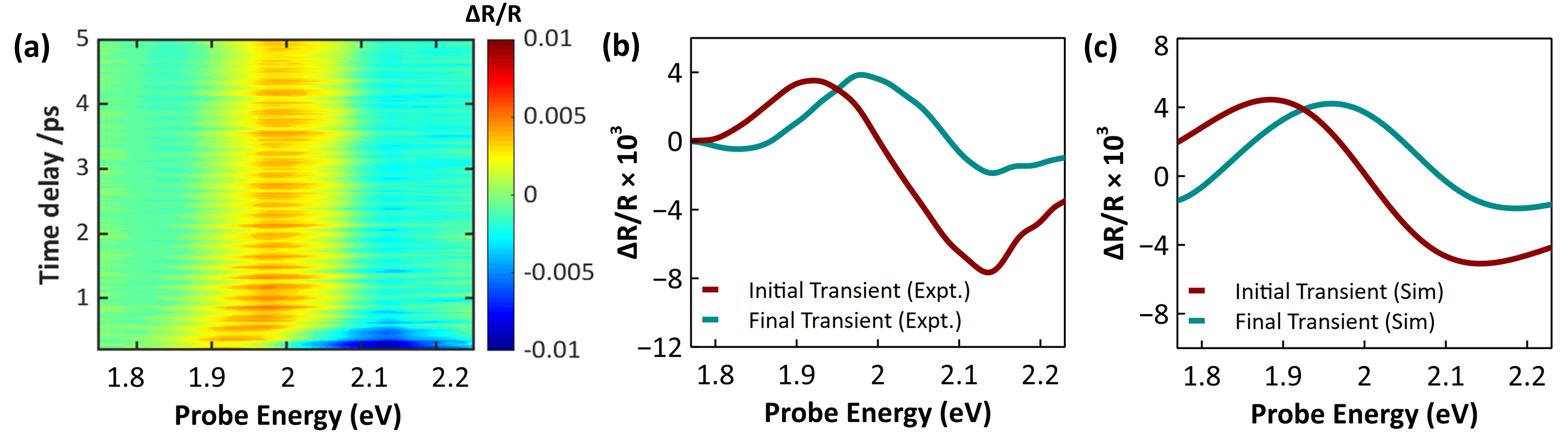}
\centering
 \caption  {(a) Transient reflectivity signal as a function of time delay and probe energy following photoexcitation with a broadband 14 fs pulse centered around 2 eV and pump fluence of 1.24 mJ/cm$^2$. (b) Initial (200 fs) and final transient (2 ps) reflectivity spectra of ZrSiS. (c) Simulated reflectivity spectra based on Drude-lorentz oscillator model show good agreement with the experimental spectral line shapes.
 Details of the simulations and fitted parameters are provided in the supplemental material Sec. III . }
  \label{Fig1}
\end{figure*}


Coupling electronic states through ultrafast optical pulses is an elegant approach to drastically change macroscopic (for example, electrical, or optical) properties of topological materials out-of-equilibrium~\cite{de2021colloquium,wang2013observation,schultze2013controlling}.
Ultrafast changes can be detected in a pump-probe approach, where a high frequency external field, or pump, is used to drive the system while a subsequent weak probe field is used to detect the induced changes~\cite{orenstein2012quantum,weber2021ultrafast,rosker1988femtosecond}.
For example, ultrafast optical pulses have significantly altered the structure of Dirac or Weyl nodes by shifting the band energy, splitting, merging or even opening a gap~\cite{bucciantini2017emergent,hubener2017creating,ebihara2016chiral,chan2016type,yan2016tunable,kibis2017all}.
In addition to this fundamental physics, the out-of-equilibrium responses of Dirac and Weyl semimetals offer technological advantages, including optical switching in mid-infrared lasers\cite{zhu2017robust}, broadband infrared photodetectors\cite{yang2019ultraviolet,liu2020semimetals}, as an efficient THz source via non-purturbative nonlinear effects\cite{cheng2020efficient,almutairi2020four}, and in high frequency opto-electronic devices.\cite{zhu2017robust,meng2018three,chan2017photocurrents,schoop2018chemical}


While there are studies on the ultrafast dynamics of ZrSiX NLSMs where the transient carrier relaxation mechanisms have been discussed\cite{weber2017similar,kirby2020transient,weber2018directly}, direct manipulation of the electronic band structure, relaxation times and mass enhancement via external electromagnetic fields has remained elusive. 
Time-resolved angle-resolved photoemission (tr-ARPES) measurements in a NLSM (ZrSiSe) and in the quasi-two-dimensional semimetal BaNiS$_{2}$ have shown that photoexcitation with optical pulse causes band renormalization within a few hundreds of femtoseconds~\cite{gatti2020light,nilforoushan2020photoinduced}.
An all-optical control of such effects through the strength of the pump pulse provides an exciting opportunity to explore out-of-equilibrium phenomena as well as the possibility of correlation driven physics in this class of materials.


Here, we use broadband visible (1.75-2.25 eV) and near infrared (0.8-1.4 eV) transient optical reflectively measurements to show the effect of ultrafast optical excitations on the electronic bands, as well as on the electronic relaxation times in ZrSiS. 
Our results demonstrate the energy shift and mass enhancement of the 1.10 eV and 2.06 eV peaks within few hundreds of femtoseconds following photoexcitation.  
We additionally observe a decrease in the electronic relaxation rate with pump fluence, in contrast to the observed coherent phonon (A$_{1g}$) lifetime that remains independent. 
These findings are well captured within a tight-binding model, where the renormalization of the electronic kinetic energy due the ultrafast pump pulse leads to the observed changes of reflectivity.
Our study combines an out-of equilibrium approach with a table-top, all optical experimental platform to demonstrate that band renormalization in ZrSiS is inherently an ultrafast effect that originates from the modification of the electronic spectrum due to the pump field.


 \begin{figure*}[ht]
\includegraphics[width=4.3in]
{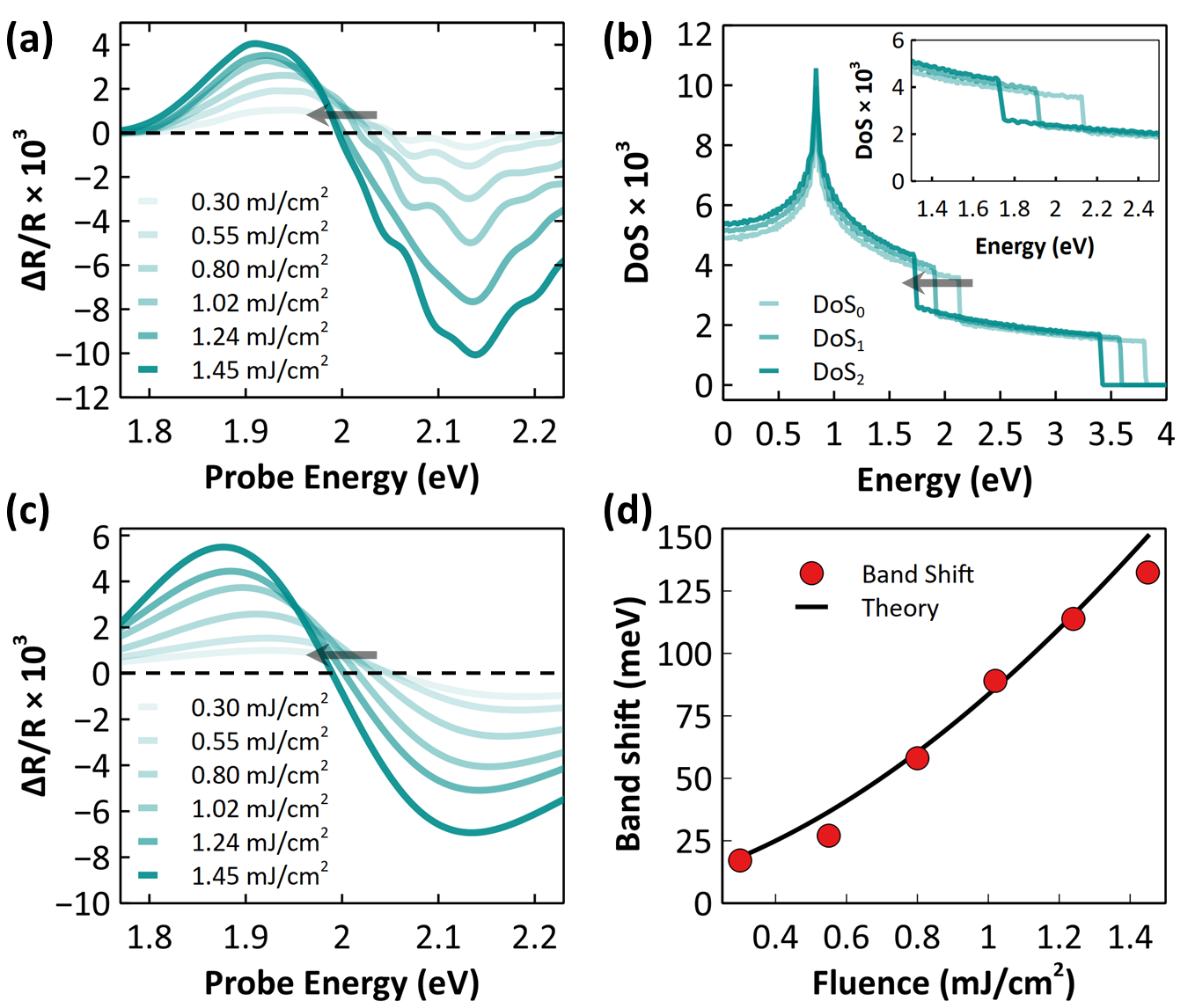}
\centering
 \caption  {(a) Fluence dependence of the initial transient spectrum. The black arrow indicates the red-shift in zero crossing energies as a function of fluence. 
 (b) Density of states obtained from the tight-binding model as a function of pump strength, DoS$_0$ to DoS$_2$ are with increasing pump fluence. 
 (c) Simulated differential reflectivity spectra as a function of fluence reproduced the experimental spectral shapes and shift in zero crossing energies. (d) The amount of band shift overlayed with the calculated band shift obtained from the tight binding model.}
  \label{Fig2}
\end{figure*}

\section{Signature of Ultrafast Band Renormalization}

First, we measured the transient reflectivity spectra from a single crystal of ZrSiS following photoexcitation using a visible broadband (1.75-2.25 eV) $14$-fs laser pulse centered around 2~eV, see Fig.~\ref{Fig1}(a) (see supplemental material sec. I and II for material characterizations and experimental details, respectively).
Initially, the differential reflectivity shows a negative change at high energies, along with a positive change at low energies. 
After $300$-$400$ fs time delay, these features change to a broad positive peak with slightly negative tails at either side, see Fig.~\ref{Fig1}(b). 
For time delays $>$500 fs, we observe no significant change in the spectral shape within the time window of our experiments.

Using the Drude-Lorentz model we fit the differential reflectivity spectra~\cite{kuzmenko2005kramers}, see Fig.~\ref{Fig1}(b) and (c), and find that the Lorentzian peak at 2.06 eV is red-shifted by 113.8 meV, while in the final transient state it recovers back to its equilibrium energy.
To understand the origin of the observed spectral features, we use a tight-binding model~\cite{Rudenko18} and calculate the transient dynamics using an out-of-equilibrium formalism~\cite{Freericks09} (see  supplemental material sec. VIII  for more details).
We find that renomalization of the kinetic energy of electrons accounts for both the shift in resonance frequency and the renormalization of the mass in the initial transient spectra around 2~eV. 
We note that the final spectral feature appears within 300 fs, which is of the same order as the electron relaxation time constant ($\sim$200 fs) reported before for ZrSiS~\cite{kirby2020transient,weber2018directly,weber2017similar}.
Therefore, the recovery of the band structure back to its equilibrium is primarily attributed to electronic effects.

\section{Fluence Dependence of the Band Renormalization}

In order to confirm that the observed spectral feature is an ultrafast electronic effect, we performed an excitation density dependence study, where the pump fluence has been tuned between 0.3 to 1.45 mJ/cm$^{2}$. 
The fluence dependence of the initial spectra is shown in Fig.\ref{Fig2}(a). 
Interestingly, we observe a red-shift of the reflectivity spectra as a function of fluence and, in particular, a shift of the zero-energy crossing. 
In contrast, all the final spectral features look identical when they are normalized (See supplemental material sec. IV). 
This suggests that ultrafast pump field affects the electronic structure only within few hundreds of femtosecond following photoexcitation.

 Out-of-equilibrium treatment of the tight-binding model (see  supplemental material sec. VIII  for more details) shows that the observed spectral shift is induced by ultrafast processes that modify the density of states according to the strength of the pump, see Fig.\ref{Fig2}~(b).
 The simulated initial transient spectra, shown in Figure \ref{Fig2}~(c), have the expected behaviour with respect to fluence: we observe a shift in zero-energy crossing of the differential reflectivity spectra, as indicated by the black arrows in Figure \ref{Fig2}(a) and (c).
Specifically, we find that the position of the Lorentzian peak centered at 2.06 eV shifts to lower energies and follows a quadratic behavior as a function of pump fluence, see Figure \ref{Fig2}(d).
In addition, the effective mass associated to this peak is shown to renormalize depending on the driving strength. 
Quasiparticle mass enhancement has been observed before in the presence of high magnetic fields~\cite{pezzini2018unconventional}. 
Therefore, our observation demonstrates that ZrSiS could be a platform to study correlation driven physics where we have the ability to tune the effective mass as a function of pump field strength; a promising avenue towards engineering non-equilibrium novel phases in this material using ultrafast optical pulses.

\section{Effect on Electronic Relaxation}

\begin{figure}[ht]
\includegraphics[width=2.7in]
{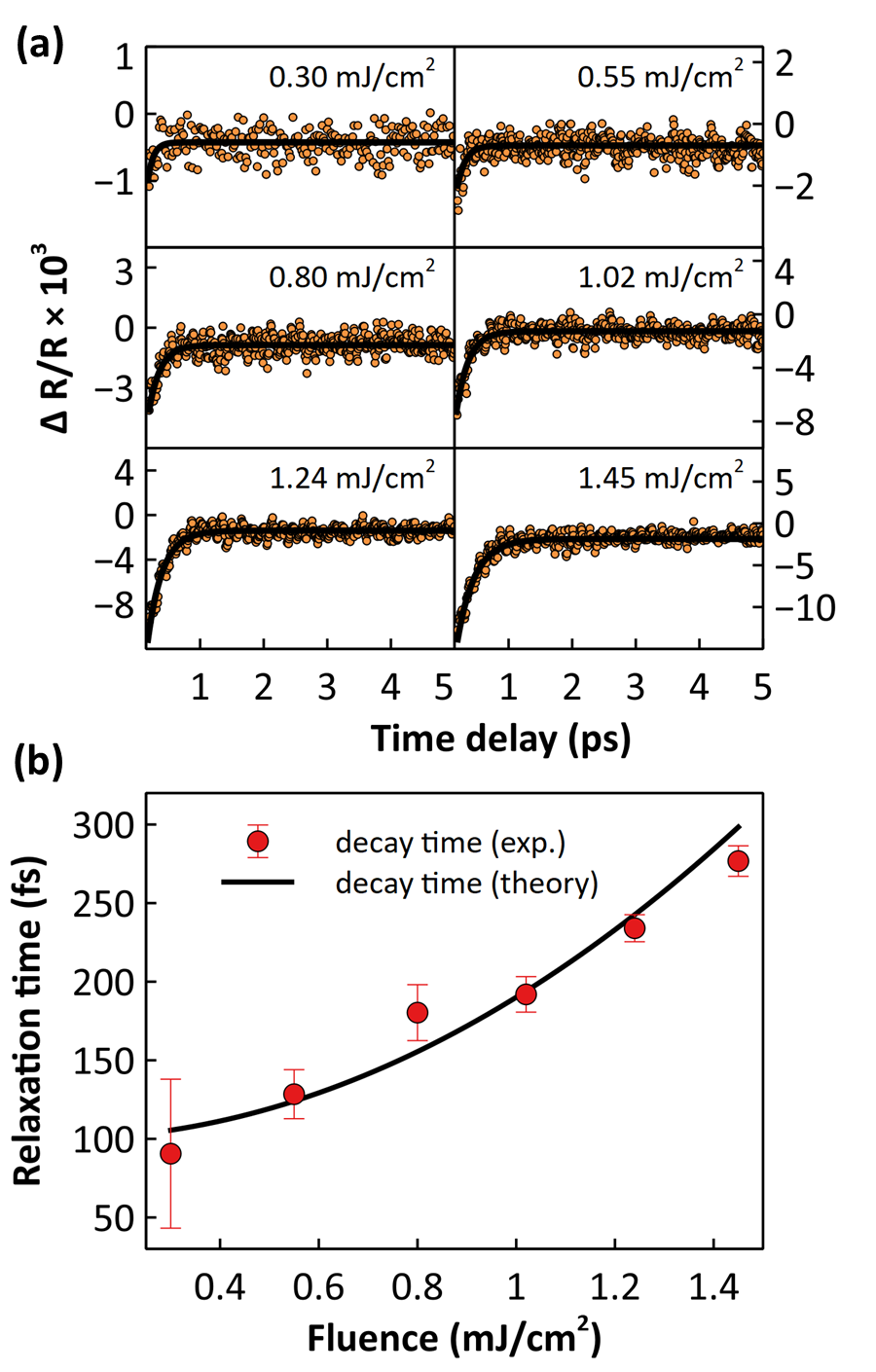}
\centering
 \caption  {(a) Fluence dependence decay of the negative reflectivity feature centered around 2.1 eV. The orange circles are the experimental data while the solid black lines are fits with a mono exponential decay function (b) Experimental relaxation time overlayed with the relaxation time obtained from out-of-equilibrium formalism}
  \label{Fig3}
\end{figure}

Next, we study the time evolution of the transient reflectivity signal with pump fluences 0.30 mJ/cm$^{2}$ to 1.45 mJ/cm$^{2}$ and extract the relaxation time by observing the decay of the negative reflectivity feature centered at 2.1 eV.
The data can be well fitted with a mono exponential decay, where the fitted relaxation time ranges between $90$ fs to $280$ fs depending on the pump fluence, see Figure \ref{Fig3}. 
We note that the energy zero-cross of the differential reflectivity spectral feature as a function of excitation density follows a similar trend to the decay of the negative peak centered around 2.1 eV (see supplemental material sec. V). 
Since the recovery of the band back to its equilibrium and electronic decay is simultaneous, they must be induced by the same ultrafast process.
Indeed, such relaxation dynamics are well captured within the tight-binding model, where the electronic recovery time depends on the pump's fluence, see Fig.\ref{Fig3}b


 \begin{figure*}[t]
\includegraphics[width=5in]
{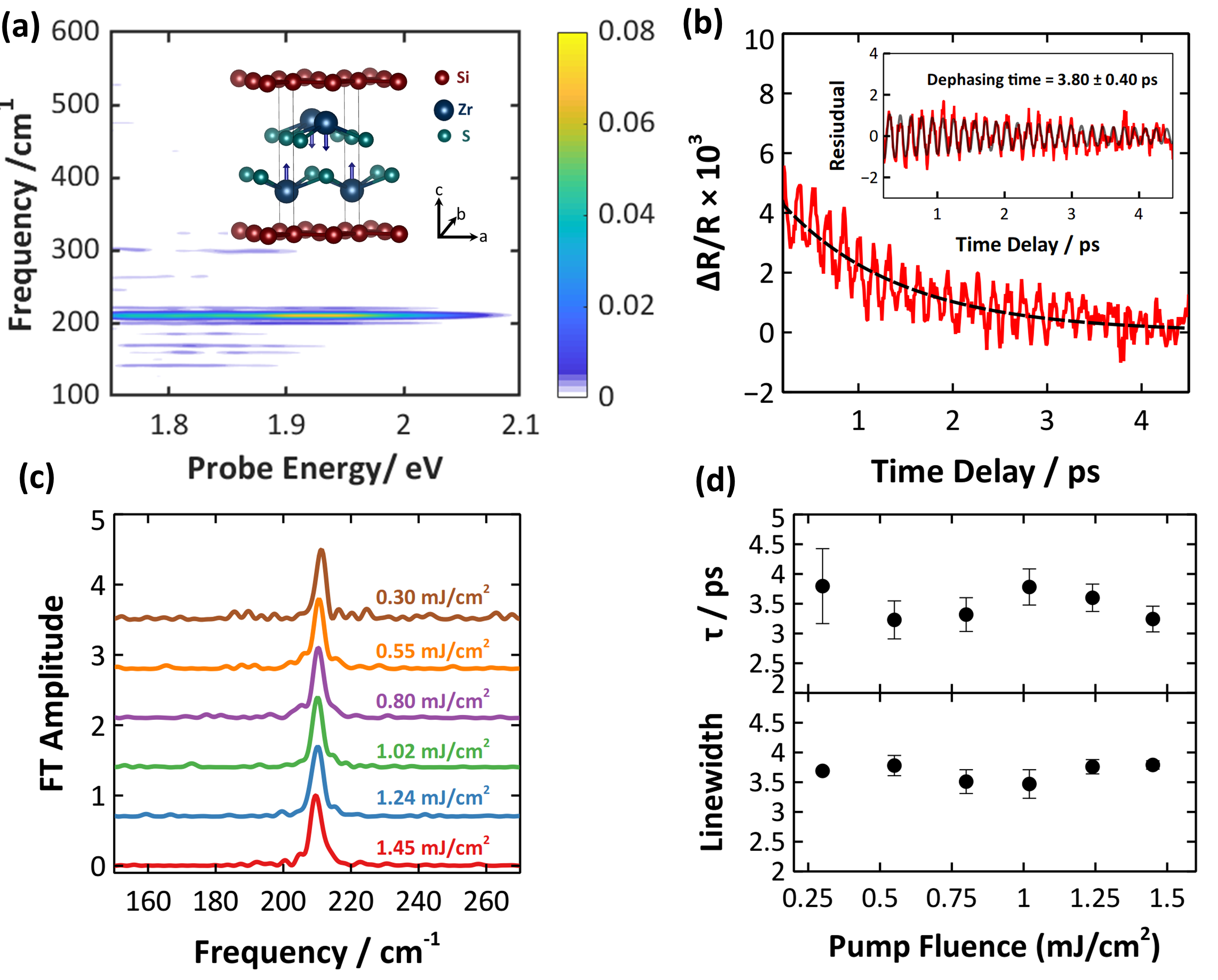}
\centering
 \caption  {(a) Fourier transformed data as a function of probe energy in the visible region showing a dominant frequency centered around 210 cm$^{-1}$ (b) Linecut at 1.9 eV of the transient differential reflectivity data as a function of time delay (red line). The dotted black line is the exponential fit. The inset shows the residual with a period of 159 fs and dephasing time constant of 3.8 ps. (c) The lattice vibration of ZrSiS of A1g phonon mode. (d) The Fourier transformed frequency as a function of pump fluence. (e) Phonon dephasing rate and linewidth as a function of pump fluence}
  \label{Fig4}
\end{figure*}

Additional transient reflectively measurements with 2eV pump and broadband near infrared (NIR) (0.8-1.4 eV) probe in the NIR region shows similar enhancement of the electronic relaxation time as a function of fluence (see supplemental material sec. VI).
The density of states obtained from the tight binding model shows that the resonant peak position near 1eV does not change as a function of pump fluence (see Fig.\ref{Fig2}b), consistent with the experimental observations (see supplemental material sec. VI); in contrast, its effective mass renormalizes depending on fluence (see supplemental material Table IV).
Therefore, the change in the electronic relaxation time is attributed to the renormalization of the effective mass, which additionally alters electronic band structure near 2eV as discussed above.
The change in electronic relaxation time due to external perturbation is in contrast to the previous measurements, where the electronic relaxation rate is independent of several experimental parameters including pump fluence, temperature, and magnetic fields in ZrSiS~\cite{kirby2020transient,weber2017similar}. 
Therefore, our study shows the first experimental evidence of affecting the electronic band structure and subsequent relaxation in a topological semimetal using ultrashort optical excitation.
While the theory predicts that the observed trend in relaxation rate as a function of fluence is an electronic effect, we note that this trend could also be due to the hot-phonon bottleneck effect where the decay of the optical phonons controls the electronic relaxation. 
In the following we confirm the validity of the theory by observing coherent phonon dynamics.


\section{Coherent Phonon Dynamics}

Along with the transient change in the reflectivity, we additionally observe strong modulation of the transient signal as a function of time in the visible pump-probe measurements (near 2 eV), see Figure~\ref{Fig4}. 
This modulation originates from the changes in the permittivity of the material due to atomic vibrations of the coherent Raman active phonon mode. 
Fourier transformation of the oscillatory signal shows a frequency at around 210 cm$^{-1}$ with a strong vibration amplitude between 1.75-2.10 eV, see Figure \ref{Fig4}(a).
This mode is a symmetric A$_{1g}$ vibration, which is an out-of-plane motion of Zr and S along c direction of the ZrSiS lattice as shown in the inset of Figure \ref{Fig4}(a) and reported previously by Raman measurements~\cite{zhou2017lattice,singha2018probing}. 
We attribute the observed coherent oscillation to the inter-band coupling between electronic modes at $\Gamma$ point~\cite{zhou2017lattice} and symmetric lattice vibration in ZrSiS; such symmetric coherent vibrations have also been observed before in other topological semimetals~\cite{kirby2021signature,weber2018using}.

Fitting the oscillatory component with a sinusoidal wave multiplied with an exponential decay function, the extracted lifetime of the coherent phonon motion is 3.80$\pm$0.40 ps (see supplemental material sec. VII). 
Surprisingly, this is relatively long compared to the coherent vibration observed in similar topological material ZrSiTe, where the oscillation diminished within 1 ps~\cite{kirby2021signature}.
In our case, the observed coherent phonon lifetime and the linewidth are independent of fluence, with an almost constant lifetime of 3.5$\pm$0.3 ps, see Fig.~\ref{Fig4}(d).
In addition, the coherent phonon lifetime is more than an order of magnitude higher than the observed electronic relaxation time, i.e., the recovery of the bands. 
These observations indicate that phononic contributions to the total change of relaxation rate are negligible and rule out the possibility of phonon-bottleneck formation: a process by which carrier lifetime is increased based on phonon relaxation. 
Our results is consistent with the predicted weak electron-phonon interactions in ZrSiS~\cite{rudenko2020electron}.

\section{Conclusions}

Combining non-equilibrium formalism with broadband optical reflectivity measurements in the visible and NIR regions, we provide evidence of band renormalization in a topological nodal-line semimetal ZrSiS. 
Our fluence-dependence study shows a red-shift in resonant frequencies and mass enhancement, along with an increase of the electronic relaxation time scales. 
These observations are inherently an ultrafast effect, as supported by a tight-binding model and by the behavior of long lived coherent phonons. 
Our study demonstrate the ability of all-optical manipulation of band structure properties, including mass enhancement and electronic relaxation, of ZrSiS, therefore, offering a unique platform to understand light driven photo-induced phases in topological materials.

\vspace{0.7cm}

\begin{acknowledgments}

G.D.S. acknowledges support by NSF through the Princeton Center for Complex Materials, NSF-MRSEC (grant
number DMR-2011750). Work by I.P. and P.N. is partially supported by the Quantum Science Center (QSC), a National Quantum Information Science Research Center of the U.S. Department of Energy (DOE). 
I.P. is supported by the Swiss National Science Foundation (SNSF) under project ID P2EZP2\_199848. 
P.N. acknowledges support from a CIFAR BSE Catalyst Program that has facilitated interactions and collaborations with G.D.S. L.S. acknowledges support by NSF through the Princeton Center for Complex Materials, a Materials Research Science and Engineering Center DMR-2011750, by Princeton University through the Princeton Catalysis Initiative, by the Gordon and Betty Moore Foundation's EPIQS initiative through Grant GBMF9064, by the David and Lucile Packard foundation and the Sloan foundation.

\end{acknowledgments}

\nocite{*}

\bibliography{apssamp}

\end{document}